# Density functional calculations for stability of Titanium-doped Gold clusters Au$_n$Ti (n=2-16)

Ming-Xing Chen[1,2], Xiao-Hong Yan[1+]

[1]College of Science, Nanjing University of Aeronautics and Astronautics, Jiangsu 210016, China

[2]Department of Physics, Xiangtan University, Hunan 411105, China

Abstract

The stability and structures of Titanium-doped Gold clusters Au$_n$Ti (n=2-16) are studied under the relativistic all-electron density-functional calculations. It is found that the most stable structures of Au$_n$Ti clusters with n=2-7 are planar. A structural transition of Au$_n$Ti clusters from two-dimensional to three-dimensional geometry occurs at $n$=8, while the Au$_n$Ti ($n$=12-16) prefer gold cage structure with Ti atom locating at the center. The size-dependence of cluster properties such as binding energy, HOMO–LUMO gaps, ionization potentials, and electron affinities have been calculated and analyzed. The Au$_{14}$Ti cluster is found to have special stability, which may be due to the electron shell effects. Further calculations are performed to study cluster-cluster interaction between two Au$_{14}$Ti clusters.

[+] Electronic mail: xhyan@nuaa.edu.cn

# I. INTRODUCTION

Transition metal (TM)-noble metal alloy clusters constitute an exciting field of research due to that they present a number of structures and properties, which are different from those of corresponding pure metals clusters and can be controlled by size and concentration.[1-10] Since a few highly stable clusters having an gold-caged structure with large highest occupied molecular orbital (HOMO)-lowest unoccupied molecular orbital (LUMO) gaps have been reported,[11-14] TM-Au binary clusters have attracted much attention.

Experimentally, Neukermans et al investigated the stability of cationic clusters $Au_nM^+$ with M from Sc to Ni, by photofragmentation experiments.[15,16] The observed intensity shown an enhanced abundance for specific cluster sizes and could be discussed on the basis of the jellium model. Photoelectron-spectroscopy studies show that TM impurity atom has a significant effect on the electronic properties of gold clusters.[17,18] More recently, combined with theoretical work, Li's experiment has demonstrated that for planar $MAu_6^-$ and $MAu_6$ (M =Ti, V, Cr) clusters, the transition metal dopant atom and the $Au_6$ ring have a unique host-guest interaction, giving rise to the atomic magnetism.[19]

Motivated by the experimental progresses, theoretical calculations of geometrical and electronic structures of the TM doped Au clusters have been performed. The structure and electronic properties of bimetallic $Au_nZn$ (n≤6) clusters have been studied with DFT calculations.[20] $Au_5Zn^+$ and $Au_4Zn$ clusters were found to have a magic number of delocalized electrons for two-dimensional systems. It is also found doping $Au_4$ cluster with Ti or Zr gives rise a magic cluster which can serve as cluster-assembled materials.[21] First-principle calculations of $Au_nM^+$ (M=Sc, Ti, V, Mn Fe, Au; $n$≤9) indicate that the local magnetic moment of the dopant atom exhibits a pronounced odd-even oscillation with the number of Au atoms, and decreases when the cluster size increases.[22] Calculations on a single 3, 4d impurity encapsulated in an icosahedral $Au_{12}$ cage show that the interaction between dopant atom and host atoms play an important role in energetics and local spin magnetic moment.[23] Based on DFT study on geometric and electronic properties of $Au_nM$ (M=Ni, Pd, Pt; $n$≤7), it is suggested that the catalytic properties will be improved with the doping of transition metal.[24]

It is well known that size-dependent features of gold clusters, such as stability and electronic properties are successfully explained by electronic shell model. However, the stability of typical TM cluster can often be discussed in terms of shells of atoms, relating the number of atoms needed to form a compact symmetric structure to an enhanced stability. In order to understand the interplay between electronic and geometric effects in bimetallic clusters, in this paper, using the first-principles method based on density-functional theory, we study the size-dependent growth behavior and electronic properties of $Au_nTi$ ($n$=2–16) clusters. In Sec. II, a brief description of the computational methods is given. Results are discussed in Sec. III. The ground-states, stabilities, and electronic properties of $Au_nTi$ ($n$=2–16) clusters have been investigated. Section IV contains the summary of our conclusions.

**II. COMPUTATIONAL METHODS**

Our calculations on the geometric and electronic structures of $Au_nTi$ ($n$=2-16) based upon the spin-polarized relativistic all-electron density functional theory (DFT) using the $DMol^3$ package.[25] The electron density functional is treated by the generalized gradient approximation (GGA) with the exchange-correlation potential parametrized by Perdew and Wang (PW91).[26] Geometry optimizations are performed by the Broyden Fletcher Goldfarb Shanno (BFGS) algorithm. Self-consistent-field procedures are done with a convergence criterion of $10^{-3}$ a.u. on the gradient and displacement, and $10^{-5}$ a.u. on the total energy and electron density. To check the validity of the computational method used for the investigation of the $Au_nTi$ clusters, we first perform the calculation on $Au_2$ and $Ti_2$ dimers. The computed values of bond length for $Au_2$ is 2.49Å and for $Ti_2$ is 1.97Å. These results are well consistent with the experimental data of 2.47Å[27] and 1.97Å[28] for $Au_2$ and $Ti_2$, respectively. This indicates that our approach provides an efficient way to study small clusters.

**III. RESULTS AND DISCUSSION**

**A. Atomic structures**

We have considered a number of possible structural candidates for each size of $Au_nTi$ clusters to search the lowest-energy configurations. The ground-state geometries of $Au_nTi$ ($n$=2-16) clusters and several typical stable isomers are presented in Figure 1, and 2, where the impurity atom Ti is shown in a white sphere. The differences of the total binding energies between an isomer and the lowest-energy structure are also given.

Figure 1 shows the most stable structures of $Au_nTi$ ($n$ =2–7) clusters. For $Au_2Ti$, the lowest-energy structure is a linear chain in which Ti atom is at the center. A low-lying isomer almost linear is found be 0.01 eV higher in total energy, in which Ti atom is at the apex and the Au–Au distance is significantly longer than the Au dimer bond length. For small clusters with 3≤$n$≤7, $Au_nTi$ adopt planar structures as their lowest-energy geometries. $Au_3Ti$ has $C_{2v}$ rhombus ground state structures. In the case of $Au_4Ti$, the most stable structure is planar trapezoidal shaped ($C_{2v}$) with impurities centered in the middle of bottom lateral. A planar hexagon with the impurity atoms located at the center sites is found to be the lowest energy structure for $Au_6Ti$ clusters, which can be derived from the distorted pentagonal isomers of $Au_5Ti$. While the most stable structure of $Au_7Ti$ can be obtained by adding a gold atom to one side of $Au_6Ti$. The ground state structures for $Au_4Ti$ and $Au_6Ti$ accord with the existing DFT results.[19,21]

The first three-dimensional (3D) structure occurs at $n$=8. Two planar structures constructed by adding two gold atoms symmetrically on the sides of $Au_6Ti$ cluster are found to be 0.292 and 0.760 eV higher in energy. Compact geometry (8$c$) is found to be a metastable structure and 0.742 eV above the most stable structure. For n=9, several isomers are studied. The most stable structure with $D_{3h}$ symmetry can be derived from 8$a$ by capping a gold atom on the top site of Ti in $Au_8Ti$. A 3D structure with $C_{3v}$ symmetry and two planar isomers are higher than the present ground-state structure by 0.123, 0.282, and 0.898 eV respectively. For $Au_{10}Ti$, the most stable structure is an uncompleted icosahedron with $C_{2v}$ symmetry. The lowest-energy structure of $Au_{11}Ti$ can be obtained by capping a gold atom on the top of $Au_{10}Ti$ with 10$b$ structure.

The structural transition takes place at $n$=12. The ground-state geometry of $Au_{12}Ti$ cluster is an fcc close-packed structure with the Ti atom in the cage center. It should be noted that Ti bulk has higher melting temperature than Au. According to theoretical models of surface segregation in binary systems the larger atom or constituent with lower melting temperature tend to segregate at the surface. Our result on Ti segregation is, therefore, in agreement with these theories. Further more, in our calculations the Ti–Au bond (2.387 eV) is stronger than the Au–Au bond (2.150 eV) for corresponding diatomic system. So the Ti atom prefers to locate in the center in order to improve the stability of system. Other low-energy isomers such as 12b ($C_{2v}$) and 12c ($D_{6d}$) are higher in energy by $\triangle E$ =0.279 and 0.522 eV respectively. It should be mentioned that previous studies shows that the ground state structures for all of XAu$_{12}$ (X=W and Mo)[11,12] and MAu$_{12}^-$ (M=V, Nb, Ta)[13] cluster are a perfect TM-centered icosahedron. Here the icosahedral structure with Ti in the center is found to be not favorable by 0.604 eV higher than *fcc* structure. The lowest-energy structure of Au$_{13}$Ti can be constructed by capping an additional Au atom on the Au$_{12}$Ti cluster. The low-energy isomers such as 13b ($C_s$), 13c ($C_{2v}$), and 13d ($C_{6v}$) have higher total energies than the 13a by $\triangle E$=0.059, 0.217, and 0.235 eV respectively.

The lowest-energy structure of Au$_{14}$Ti is similar to those of Au$_{14}$Zr and Au$_{14}$Hf,[13] in which 14 gold atoms form a hollow cage with $D_{2d}$ symmetry, while the TM atom is encapsulated in the gold cage. Two Au-caged structures with $C_{2v}$ symmetry (14b) and $D_{6d}$ symmetry (14c) are in competition with the ground-state geometry, and the relative energy are merely 0.035 and 0.044 eV. A metastable isomer (14d) with two Au atoms capping on the icosahedron is above the lowest energy geometry by 0.112 eV. For Au$_{15}$Ti, the lowest energy structure can be viewed as a consequence of replacing the top Au atom of Au$_{14}$Ti with two Au atoms. Three isomers are slightly higher in energy with $\triangle E$=0.040, 0.101, and 0.204 eV respectively. The ground state structure for Au$_{16}$Ti is also a continuation of the structural pattern of *n*=14 (14a) and *n*=15 (15c). The structure with $C_s$ symmetry can be yield as an additional atom being capped on Au$_{15}$Ti cluster. It is more stable over another isomer (16d) with similar geometry ($C_{2v}$) by 0.552 eV.

**B. Stabilities and electronic properties**

The stability of these clusters can be discussed on the basis of the binding energy per atom ($E_b$), the second-order energy difference ($\Delta_2 E$), $E_b$ (Au$_n$Ti) = [E (Ti) + *n* E (Au) - E (Au$_n$Ti)] / (*n*+1), $\Delta_2 E$ (Au$_n$Ti)= E (Au$_{n-1}$Ti) + E (Au$_{n+1}$Ti) − 2 E (Au$_n$Ti)], where E is the total energy of the system. As shown in Fig. 2, the binding energy increases monotonically as the size of clusters increases. The little peak at n=14 shows the Au$_{14}$Ti cluster has a higher stability. For $\Delta_2 E$, being a sensitive quantity that characterizes relative stability of atomic clusters as a function of cluster size, show odd-even staggering at the range of n=2-6 and n=10-16. Notice peaks are found at *n*=4, 12, and 14, indicating that these clusters should be more stable than the clusters with neighboring sizes. Our calculations on the stability of $Au_4Ti$ show a good agreement with previous theoretical calculations, which $Au_4Ti$ was predicted to be a magic cluster and can serve as building blocks for cluster-assembled materials.[21] For the n=12 is the transition size for the formation of Au cage, thus the enhanced stability of $Au_{12}Ti$ may be attributed to the geometrical effects. For the four valence electrons of Ti along with the 14 valence *s* electrons of Au atoms in $Au_{14}Ti$ result in a total of 18 electrons, one could suspect that the enhance stability of $Au_{14}Ti$ may be due to the closure of electron shell.

The vertical ionization potential (VIP) is a useful quantity for determining the stability of clusters. It is defined as the total-energy difference of the neutral cluster and the ionized cluster with the same geometry as the neutral. The VIP increases from n=3 – 6, then drastically decreases as cluster size and shows an odd-even pattern at the range of n=8-16. $Au_{14}Ti$ is found to have a significantly large ionization potential than its neighboring clusters. In addition, we have also calculated the vertical electron affinities (EA) by assuming the geometry for the charged cluster to be the same as the neutral one. The EA exhibit an approximate odd-even oscillation from n=5 up to 16. Noticeable peaks are found at n=3, 6 and 13, and dips are found at *n*=4 and 14, which indicate that closed electron shell occurred at $Au_4Ti$ and $Au_{14}Ti$. The high EA for $Au_6Ti$ may be due to its peculiar structure. For $Au_{13}Ti$, the high electron affinity can be understood that it has 17 valence electrons, one electron less than the 18 electrons closed shell. Consequently, $Au_{14}Ti$ clusters may have a low electron affinity because of the closure of electron shell as addition of one gold atom to $Au_{13}Ti$ cluster.

The HOMO-LUMO energy gap is another useful quantity for examining the stability of clusters. As is known, systems with larger gap are less reactive. In Fig. 3, one can find that $Au_4Ti$, $Au_7Ti$, $Au_{10}Ti$ and $Au_{12}Ti$ have a relative large gap than their neighboring clusters. Obviously, the $Au_{14}Ti$ cluster has a significantly large gap beyond 1.40 eV. The calculated values of the ionization potential, electron affinity and the HOMO-LUMO gaps for $Au_{14}Ti$ along with $\Delta_2E$ as discussed above are clearly large enough for it to qualify as magic clusters. In addition to $Au_{14}Ti$ large HOMO-LUMO gaps are found for $Au_{13}Ti^-$ (1.34 eV) and $Au_{15}Ti^+$ (1.38 eV) in our calculations, which are comparable to that of $Au_{14}Ti$. The large HOMO-LUMO gap may be one of the main reasons for the enhanced stability of $Au_{15}Ti^+$ observed by previous experiment (Ref. 15). Considering that both $Au_{13}Ti^-$ and $Au_{15}Ti^+$ clusters are valent isoelectronic with $Au_{14}Ti$, one can believe that the large gap of these clusters could be mainly due to the closure of electron shell.

**C. Cluster dimer**

It is believed that magic clusters can serve as building blocks for nanostructures. It has been recognized that the structural identity of the magic clusters in the assembly can be best retained due to shell closings of the valence electrons or geometric factors. And these clusters are usually expected to interact rather weakly *via* a van der Waals like mechanism with a very small interaction energy (~0.5 eV).[29] However, $MAu_4$ (M=Ti and Zr) cluster shows anomalous behavior.[21] Considering $MAu_4$ have an opened geometrical structure, here we calculated the interaction between two $Au_{14}Ti$ clusters to investigate the role of geometry on the interaction. Since the nonspherical shape of the $Au_{14}Ti$ cluster, there are many possibilities for the relative orientation of two $Au_{14}Ti$ clusters. Here, we just consider three configurations as indicated in Figure 5, apex-apex (a), face-face (b), and side-side (c). The average cluster-cluster distance $R$ (Å), cluster-cluster interaction energy $E_b$ (eV) and number $n$ of Au-Au bond between two clusters are summarized in Table.1. For configuration 1 (Figure 5*a*), the cluster dimer is formed by apex to apex connections of the $Au_{14}Ti$ clusters. The nearest atomic distance $R$ between the two clusters is 2.636 Å (Table. 1), which is slightly smaller than the Au-Au bond length (2.730 Å) in $Au_{14}Ti$. However, the calculated cluster-cluster interaction energy is found to be 0.746 eV, which is much smaller than 2.570 eV of Au-Au binding energy in $Au_{14}Ti$ cluster. In configuration 2 (Figure 5*b*), which is constructed by face to face, the clusters just undergo a little geometric distortion after optimization. The average bond length (2.686 Å) between two clusters is slightly larger than that of configuration 1. However, cluster-cluster interaction energy is unusually high (3.503 eV) relative to the weak interaction in configuration 1. In the case of configuration 3 (Figure 5*c*) constructed by side to side, there is significant distortion in the initial configuration during the optimization. The average bond length $R$ between two clusters is 2.684 Å and the cluster-cluster interaction (4.093 eV) is even stronger than that of configuration 2. As can be seen from Table 1, the cluster-cluster interaction is related to the number of Au-Au bond between two clusters. There is only one Au-Au bond between two clusters in configuration 1. However, in the case of configuration 2 and 3, there are much larger number of Au-Au bonds, which lead to stronger cluster-cluster interaction. Thus it may be reliable to believe that the relative orientation plays an important role in the cluster-cluster interaction. These calculations indicate that even these clusters that have closed electronic shell structure, such as $Au_{14}Ti$, still interact very strongly, and react chemically with one another to form a new species. They would certainly not retain their individual identities in the way that the fullerenes do, when used to generate a nanostructured array.

**IV. CONCLUSIONS**

In summary, we have studied the structures and electronic properties of Ti-doped Au clusters by employing density-functional theory. The stable geometrical structures of $Au_nTi$ ($n=2 - 16$) are obtained by using the *ab initio* calculations with generalized gradient approximation. The systemic and detailed analysis has been carried out in order to verify the stability of $Au_nTi$ clusters. The pronounced character common to $\triangle_2 E$, VIP, EA and the HOMO-LUMO gap at $n=14$ suggest that $Au_{14}Ti$ is a magic cluster. Further more, we have carried out calculations to study cluster-cluster interaction between two magic $Au_{14}Ti$ clusters. Whether the clusters retain their individual identities depends on the relative orientation of two clusters.

**Acknowledgements**

One of the authors (Ming-Xing Chen.) especially thanks Prof. X.G. Gong for helps. This work was supported by the National Natural Science Foundation of China (Grant No10674070), and partly by the Program for New Century Excellent Talents in University (Grant No NCET-04-0779) and the Program for Changjiang Scholars and Innovative Research Team in University (Grant No IRT0534).

Table 1. The average cluster-cluster distance $R$ (Å), cluster-cluster interaction energy $E_b$ (eV) and number $n$ of Au-Au bond between two $Au_{14}Ti$ clusters in $(Au_{14}Ti)_2$ dimer. Configuration 1, 2, and 3 correspond to the structure *a, b* and *c* in Figure. 5, respectively.

| Configuration | $R$ (Å) | $E_b$ (eV) | $n$ |
|---|---|---|---|
| 1 | 2.636 | 0.746 | 1 |
| 2 | 2.686 | 3.503 | 6 |
| 3 | 2.684 | 4.093 | 8 |

Figure 1 Equilibrium geometries of the few isomers of cationic $Au_nTi$ clusters, with n=2-11. The gold atoms are represented by dark spheres. The differences of total binding energies (in eV) of an isomer from the most favorable isomer are given below the structure for each size.

Figure 2 Equilibrium geometries of the few isomers of cationic $Au_nTi$ clusters, with n=12-16. The differences of total binding energies (in eV) of an isomer from the most favorable isomer are given below the structure for each size.

Figure 3 (a) Second-order energy differences ($\Delta_2E$) and (b) binding energy (eV) of the ground state versus number of gold atoms for $Au_nTi$ clusters.

Figure 4 (a) The vertical ionization potential (VIP), (b) electron affinities (EA), and (c) the HOMO-LUMO energy gap shown as a function of the number of Au atoms.

Figure 5. Optimized atomic structures for different configurations of cluster-cluster interaction between two $Au_{14}Ti$ clusters, (a) apex-apex, (b) face-face, and (c) side-side.

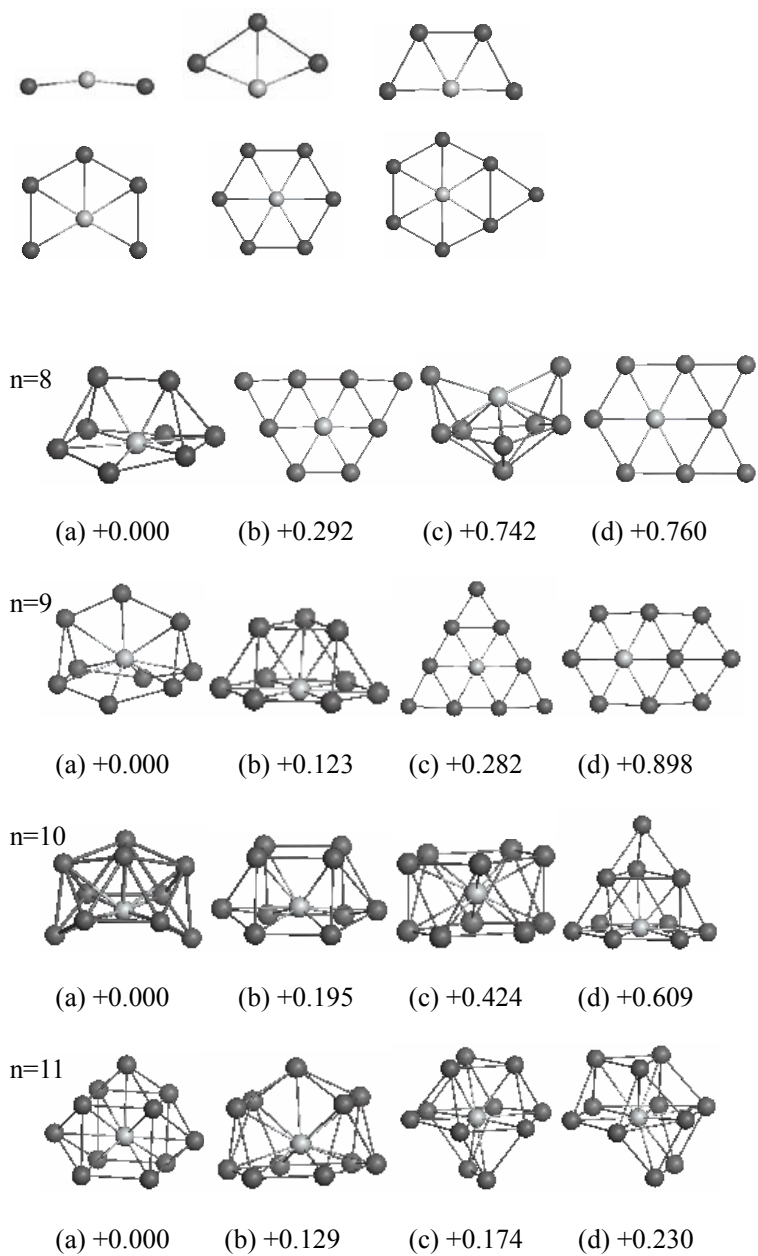

n=8
(a) +0.000   (b) +0.292   (c) +0.742   (d) +0.760

n=9
(a) +0.000   (b) +0.123   (c) +0.282   (d) +0.898

n=10
(a) +0.000   (b) +0.195   (c) +0.424   (d) +0.609

n=11
(a) +0.000   (b) +0.129   (c) +0.174   (d) +0.230

Figure 1

n=12

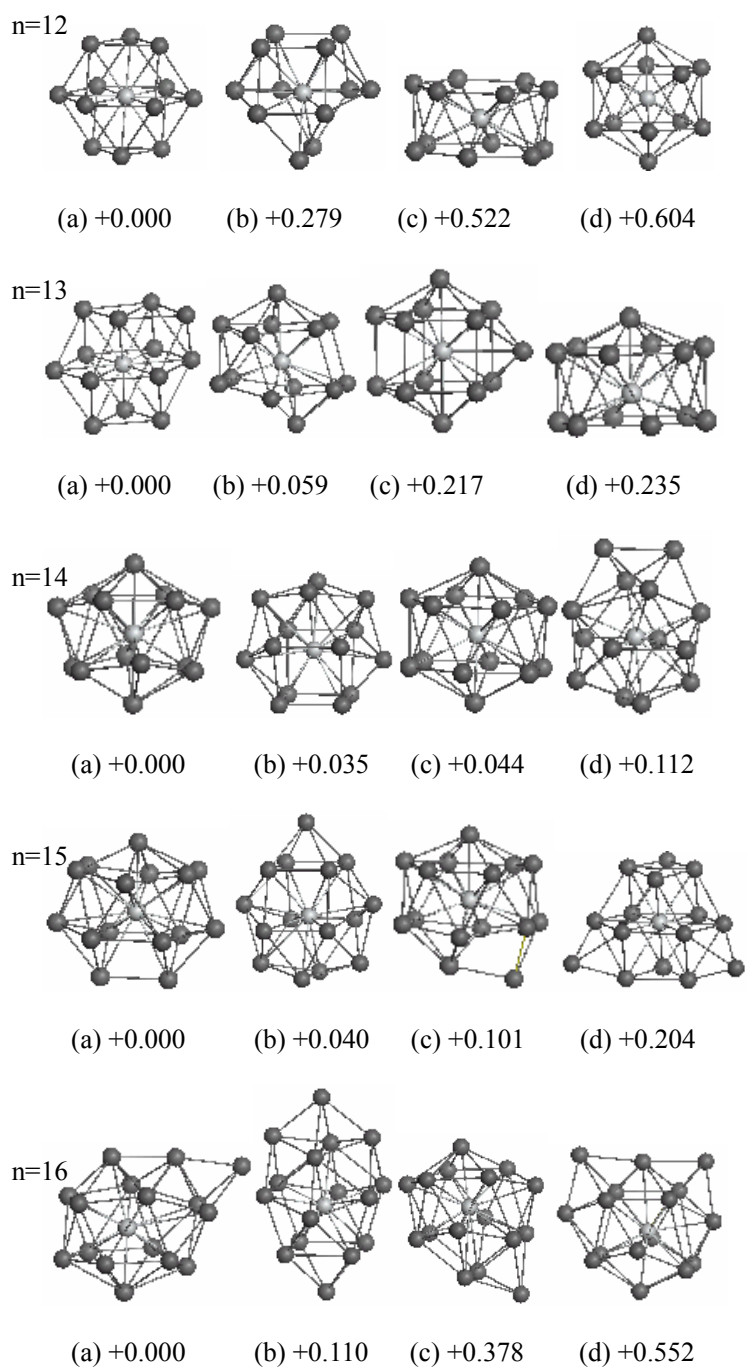

(a) +0.000    (b) +0.279    (c) +0.522    (d) +0.604

n=13

(a) +0.000    (b) +0.059    (c) +0.217    (d) +0.235

n=14

(a) +0.000    (b) +0.035    (c) +0.044    (d) +0.112

n=15

(a) +0.000    (b) +0.040    (c) +0.101    (d) +0.204

n=16

(a) +0.000    (b) +0.110    (c) +0.378    (d) +0.552

Figure 2

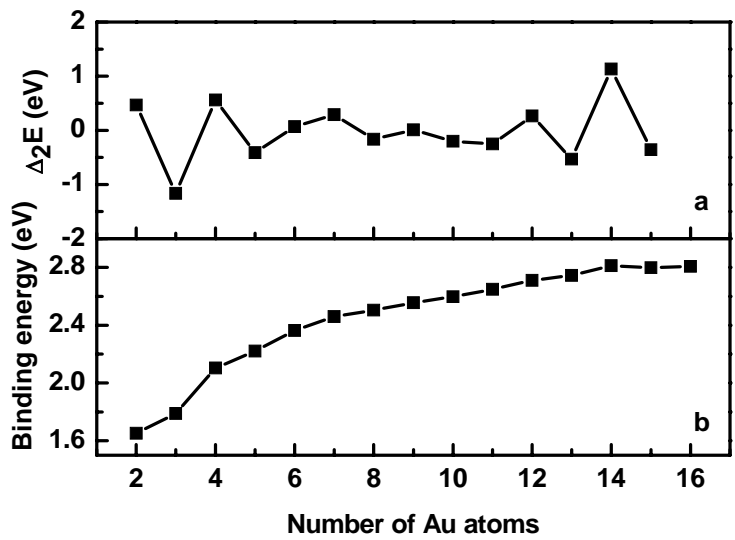

Figure 3

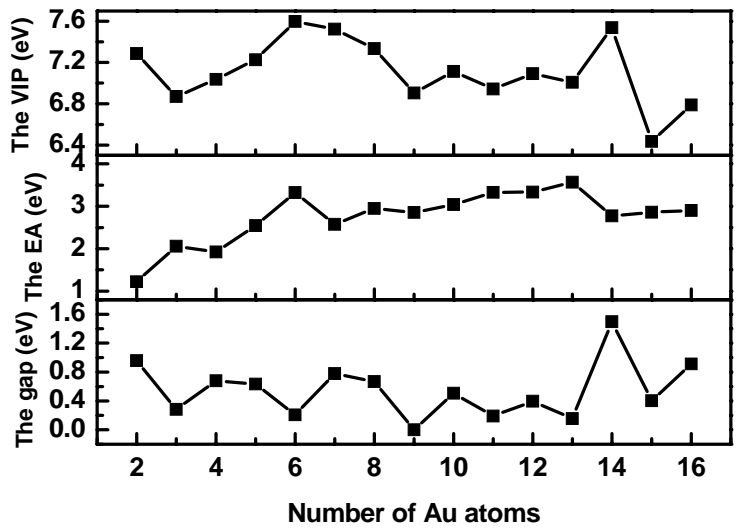

Figure 4

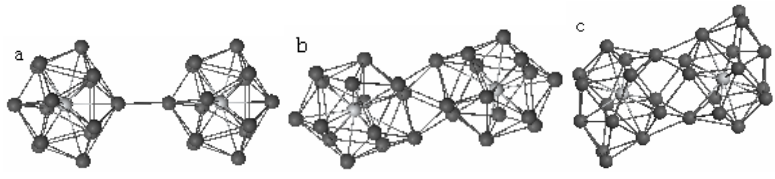

Figure 5